\documentclass[aps,pra,nofootinbib,twocolumn,floatfix,showpacs,superscriptaddress]{revtex4-2}
\usepackage[utf8]{inputenc}
\usepackage{amsmath,graphicx,epsfig,amsthm,color,bm,braket}
\usepackage{pifont,bbding,amssymb,soul,mathrsfs}
\usepackage{indentfirst}
\usepackage{times,dsfont,units}
\usepackage{appendix, comment}

\newcommand{\Tr}{\text{Tr}}

\usepackage{dcolumn}

\usepackage[hyperindex,breaklinks,colorlinks=true,citecolor=blue,urlcolor=blue]{hyperref}

\begin{document}
\title{Experimental Acceleration of Spin Transition in Nitrogen-Vacancy Center}

\author{Si-Qi Chen}
\affiliation{School of Physics, State Key Laboratory of Crystal Materials, Shandong University, Jinan 250100, China.}
\author{He Lu}%
\email{luhe@sdu.edu.cn}
\affiliation{School of Physics, State Key Laboratory of Crystal Materials, Shandong University, Jinan 250100, China.}


\begin{abstract}
Shortcuts to adiabaticity~(STA) enables fast and robust coherent control of quantum system, which has been well placed in quantum technologies. In particular, inverse engineering STA provides much more freedom for the optimization of shortcut, which alleviates the complexity for experimental realization. Here, we implement a STA technique, known as invariant-based inverse engineering, to speed up the adiabatic control of the electron triplet ground state of a single nitrogen-vacancy~(NV) center. The microwave pulses to drive inversely engineered STA are obtained with space
curve quantum control, where the evolution of spin transition is mapped to a three-dimensional closed space curve and the design of shortcut is obtained by optimization over the space curve. We demonstrate the fast and high-fidelity drive of dipole-forbidden transition between two spin sublevels of the ground state. Moreover, we demonstrate the robustness of the spin transition by introducing the detuning of driving microwave field. The acceleration and robustness is further confirmed by the comparison with two traditional Raman control schemes. Our results suggest invariant-based inverse engineering is powerful for fast and robust manipulation of NV system, and thus benefits quantum sensing and quantum computation based on the NV platform. 
\end{abstract}

\maketitle




The nitrogen-vacancy~(NV) center in diamond emerges as one of the most promising solid-state platforms for advanced quantum technologies, due to its long-lived spin states and well defined optical transitions~\cite{Jelezko2004PRL,Golter2014PRL}. The spins associated to the defects enable quantum sensing~\cite{dSchirhagl2014ARPC, Barry2020RMP}, quantum communication \cite{Bernien2013Nature,Hensen2015Nature,Humphreys2018Nature} and quantum computation~\cite{Childress2013,Pezzagna2021APR},   
where the complete and robust control of spin state is the prerequisite to demonstrate quantum advantages. {The approaches to demonstrate robust quantum control have been widely investigated, such as adiabatic quantum control~\cite{Childs2001PRA,Kral2007RMP}, single-pulse-shaping engineering~\cite{Barnes2012PRL,Daems2013PRL}, composite pulses~\cite{Brown2004PRA,Torosov2011PRL,Torosov2014PRA,Rong2015NC,Torosov2020PRRes,Dridi2020PRA,Wu2023PRA,Shi2024PRApplied}, and shortcuts to adiabaticity~\cite{Chen2010PRL,Ennio2013,Odelin2019RMP}. Among them, the adiabatic processes, such as} stimulated Raman adiabatic passage~(STIRAP)~\cite{Gaubatz1990JCA}, has been a routine technique used for robustly implementing quantum state transition in various quantum system including NV center~\cite{Golter2014PRL, ex_bohm2021PRB}, superconducting qubits~\cite{Kumar2016NC,Xu2016NC}, photon~\cite{Kuhn2002PRL}, ultracold molecules~\cite{Winkler2007PRL,Johann2008Science,Molony2014PRL}, ions~\cite{Higgins2017PRL} and optomechanics~\cite{Fedoseev2021PRL}. However, adiabatic evolution is generally time consuming~\cite{Messiah1961,Kato1950JPSJ,Sakurai2020,Albash2018RMP} which limits its application when the decoherence time is on the same scale. 

Shortcuts to adiabaticity~(STA) is a promising approach to overcome the time consuming of adiabatic control by finding fast trajectories connecting the initial and final states~\cite{Chen2010PRL,Ennio2013,Odelin2019RMP}, which has been studied in state preparation~\cite{Kolbl2019PRL,Abah2020PRL,Chen2021PRL}, manipulation~\cite{Chen2012PRA,Ruschhaupt2012NJP,Kiely2014JPB,Kiely2016JPB,Benseny2017EPJ,Li2018PRA,Angelis2020PRResearch,Vitanov2020PRA},  quantum computing~\cite{Hegade2021PRApplied} and quantum thermodynamics~\cite{Funo2017PRL,Abah2020PRResearch,Hartmann2020PRResearch}. There are several methods in STA. Counterdiabatic driving~(CD) relies on the use of auxiliary counterdiabatic fields~\cite{Berry2009JPA,Demirplak2003JPCA,Demirplak2005JPCB,Demirplak2008JCP}, and has been experimentally demonstrated~\cite{Zhang2013PRL,An2016NC,Du2016NC,Kolbl2019PRL,Wang2019PRApplied,Zhou2020PRApplied,Yin2022NC,Cheng2023PRA}. However, the explicit form of the required CD fields are generally complicated, which is not always feasible in practice~\cite{Bason2012NP,Baksic2016PRL,Zhou2017NP}. An alternative approach to accelerate adiabatic process is invariant-based inverse engineering approach~\cite{Chen2010PRL}, in which the Hamiltonian is found from a prescribed state evolution and it only requires the commutativity of invariant and the Hamiltonian at the start and the end of the evolution. Along this spirit, the invariant-based approach has been theoretically investigated in design of state transfer schemes in two-level~\cite{Ruschhaupt2012NJP,Martinez2013PRL,Kiely2014JPB}, three-level~\cite{Chen2012PRA,Kiely2014JPB,Benseny2017EPJ}, and four-level systems~\cite{Gungordu2012PRA,Herrera2014PRA,Kiely2016JPB}. As the commutativity of invariant and the Hamiltonian is not imposed at the intermediate time, it is flexible to design the state evolution with optimal control optimization, {where the deviation from the ideal evolution can be eliminated at least first order~\cite{Ruschhaupt2012NJP,Daems2013PRL,Yu2018PRA}.} 

{Recently, geometric formalism has been exploited to design robust single- and two-qubit gates~\cite{Zeng2018PRA,Zeng2018NJP,zeng2019PRA,Dridi2020PRL,Buterakos2021PRXQ,Dong2021PRXQ,Nelson2023PRA}. It has been shown that the state evolution and the corresponding control pulses is connected through a formalism called space curve quantum control~(SCQC)~\cite{Barnes2022QST}. In SCQC, the state evolution is mapped onto closed space curves, and the evolution-induced leading-order error can be eliminated by properly designed curves~\cite{zeng2019PRA,Dridi2020PRL,Barnes2022QST}. Accordingly, the driving pulses with SCQC are then determined by few parameters of curves, which alleviates the complexity in pulses engineering compared with other numerical optimal control techniques \cite{Nobauer2015PRL,Van_Damme2017PRA,Rong2015NC}. Moreover, the driving pulses have smooth profile compared to composite pulses ~\cite{Torosov2014PRA,Torosov2020PRRes,Dridi2020PRA} and dynamically corrected schemes~\cite{Kestner2013PRL,Wang2014PRB,rong2014PRL,Rong2015NC,Zhang2016PRA}. The advantages of SCQC has been studied in the noise-resilient Landau-Zener transitions~\cite{Zhuang2022Quantum}, the dipole forbidden transitions~\cite{li2022PRA} and quantum gates~\cite{Tang2023PRApplied,Yi2024RPL}.}

In this work, we demonstrate the speedup of transition between spin states that are not directly coupled, in the ground states of a single NV center. {The microwave~(MW) pulses driving spin transitions are tailored to achieve STA by invariant-based inverse engineering associated with SCQC, so-called STA-SCQC. In the presence of detuning error, the evolution of noisy spin transition is mapped to a three-dimensional closed space curve, and the design of shortcut with robustness constraint is obtained by optimization over the space curve. To give a comparison, we also perform two Raman control schemes, i.e., stimulated Raman transitions~(SRTs)~\cite{Linskens1996PRA,Bateman2010PRA} and STIRAP~\cite{Gaubatz1990JCA}, which are traditional schemes for selective transfer of populations between quantum states that are not directly coupled.} Compared to Raman control schemes, we observe enhancements of spin transition in terms of transition speed and the robustness against experimental imperfections with {STA-SCQC}. Specifically, the spin state transition with {STA-SCQC} is 6 time faster than that with STIRAP, and is able to tolerate more imperfections in the driven fields compared to SRT.

\begin{figure}[htbp]
\includegraphics[width=\linewidth]{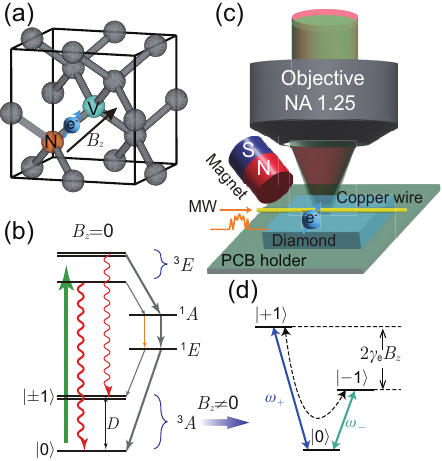}
\caption{\label{Fig:1}(a) The atomic structure of the nitrogen vacancy~(NV) center in diamond lattice. The gray, orange and green spheres represent the carbon atom, nitrogen atom and vacancy site respectively. (b) Energy-level configuration of the NV defect center. $^3A$ and $^3E$ represent triplet manifold of ground states and excited states, while $^1A$ and $^1E$ represents the singlet states. In the presence of external magnetic field along NV axis as shown in (a), the energy of states $\ket{+1}$ and $\ket{-1}$ is split by $2\gamma_eB_z$, and transition between $\ket{0}$ and $\ket{\pm1}$ can be driven by MW field with frequency $\omega_\pm$. (c) The schematic drawing of experimental setup. (d) The three-level quantum system spanned by $\ket{0}$ and $\ket{\pm1}$ in $^3A$, where $\ket{+1}$ and $\ket{-1}$ are split by 2.827~GHz.}
\end{figure}

As shown in Fig.~\ref{Fig:1}~(a), an NV center is formed by a single nitrogen~(N) atom substituting a carbon~(C) atom  and an adjacent carbon vacancy~(V). The NV center is able to capture an additional electron to form the negatively charged NV$^-$. The electrons bound to NV center have spin $S=1$, forming a triplet manifold $^3$A of states $\ket{m_s=0}$ and $\ket{m_s=\pm1}$ with a zero-field splitting of $D=2.87$~GHz as shown in Fig.~\ref{Fig:1}~(b). Hereafter, $\ket{m_s=0}$ and $\ket{m_s=\pm1}$ are denoted as $\ket{0}$ and $\ket{\pm1}$ respectively. $\ket{+1}$ and $\ket{-1}$ are split by $2\gamma_eB_z$ with $\gamma_e=2.8$~MHz/G being the gyromagnetic ratio of electron, in the presence of external magnetic field $B_z$ along the NV axis. All the spin states ($m_s=0,\pm1$) of NV center can be optically excited from the ground state to the excited state $^3E$ through spin conserving transitions~($\Delta m_s=0$) by a 532~nm laser. After being excited, optical relaxation is taken place via radiative transition~($\Delta m_s=0$) producing a broadband red photoluminescence~(PL), or through non-radiative intersystem crossing~(ISC) to the singlet states $^1E$ and $^1A$. The the non-radiative ISC is strongly spin selective, i.e., the probability of non-radiative ISC from $m_s=0$ is much smaller than that from $m_s=\pm1$~\cite{Manson2006PRB}.  This spin-selectivity of the decay process enables the polarization of electron spin states into $\ket{0}$ after a few optical pumping cycles~\cite{Robledo2011NJP}. Also, it provides a practical way to distinguish spin states $\ket{0}$ and $\ket{\pm1}$ by optically detected magnetic resonance~(ODMR) of the NV center, as the spin state $\ket{0}$ shows more PL. The NV center is hosted in a diamond substrate with less than 5~ppb N and 1.1$\%$ $^{13}$C isotope abundance, cut perpendicular to the [100] crystallographic direction. We use a home-built confocal microscope for the selective optical excitation and detection of fluorescence from single NV centers as shown in Fig.~\ref{Fig:1}~(c). The excitation light~(532~nm) from a diode laser is digitally modulated by an acousto-optic modulator~(AOM) and then reflected to an oil-immersion objective lens~(NA=1.25) by a dichroic mirror. The PL from NV center is collected by the same microscope objective, which transmits the dichroic mirror and is detected by a single-photon detector. The MW field to drive the spin transition is delivered by a copper wire with a diameter of about 20~$\mu$m, and connected to a printed circuit board~(PCB). The PCB holder is attached on a $xy$ piezo stage, which can be scanned by $70~\mu m\times 70~\mu m$ to locate a single NV center. 

In our experiment, we focus on the three-level quantum system spanned by sublevels $\ket{0}$ and  $\ket{\pm1}$ of ground state $^3A$, in a $\vee$-type configuration as shown in~Fig.~\ref{Fig:1}~(d). By applying a static magnetic field $B_z$ generated from a permanent magnet, we split the   $\ket{+1}$ and $\ket{-1}$ by 2.827~GHz, and the resonant frequencies between $\ket{0}\leftrightarrow\ket{+1}$ and $\ket{0}\leftrightarrow\ket{-1}$ are $\omega_+=4.284$~GHz and $\omega_-=1.457$~GHz. The NV center is driven by two MW fields with frequencies of $\omega_+$ and $\omega_-$ and time-dependent amplitudes of $A(t)$. In the dipole approximation and rotating-wave approximation, the Hamiltonian~($\hbar=1$) of NV system is $H(t)=\Omega(t)J_x/\sqrt{2}$, where $\Omega(t)$ is the envelope of Rabi frequency proportional to the amplitude of field of $A(t)$, and $J_\nu(\nu= x ,y ,z)$ are spin-1 angular momentum matrices. In this three-level system, the transition between $\ket{+1}$ and $\ket{-1}$ is dipole-forbidden. Our goal is to realize the dipole-forbidden transition  $\ket{+1}\to\ket{-1}$ with time dependent modulation of $A_\pm(t)$ in the framework of invariant-based inverse engineering~\cite{Chen2010PRL}.

\begin{figure*}[ht!bp]
\includegraphics[width=\linewidth]{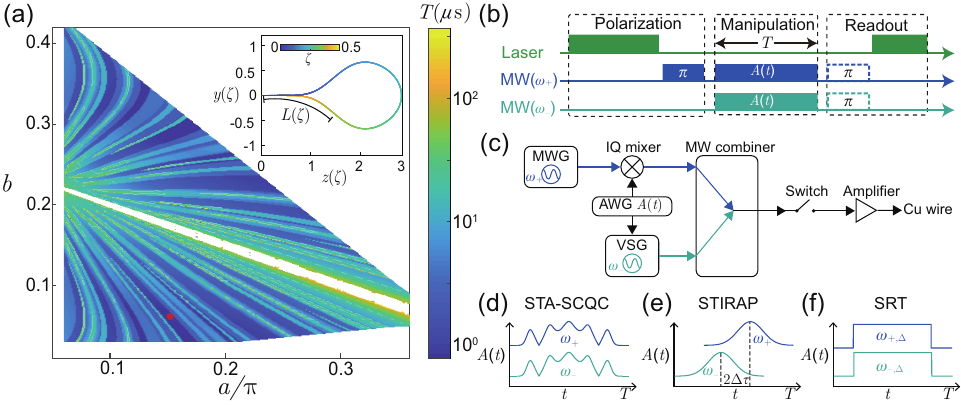}
\caption{\label{Fig:2}(a) The results of numerical simulation to find the minimal operation time $T$ to achieve the transition from $\ket{+1}$ to $\ket{-1}$. The insert is the three-dimensional curve in Eq.~\ref{Eq:spacecurve}. The red dot represents the {optimized} $T\approx0.8~\mu$s.  (b) Pulses sequence to realize spin transition. (c) The setup to generate driven MW field $A(t)$, including {STA-SCQC}~(d), STIRAP~(e) and SRT~(f).}
\end{figure*}

For Hamiltonian $H(t)$, one can construct a dynamical invariant $I(t)=\Omega_0[\sin\theta(t)\cos\beta(t)J_x+\sin\theta(t)\sin\beta(t)J_y+\cos\theta(t)J_z]$ satisfying $i\hbar\partial_tI(t)-[H(t),I(t)]=0$~\cite{Chen2012PRA,Yu2018PRA}, where 
\begin{eqnarray}
\label{Eq:theta}&\dot{\theta}=-\Omega\sin\beta,\\
\label{Eq:beta}&\dot{\beta}=-\Omega\cot\theta\cos\beta.
\end{eqnarray}
$\Omega_0$ is a constant frequency, guaranteeing the same dimension as $H(t)$. $I(t)$ is a Hermitian operator with eigenvalues and eigenvectors of  $\lambda_n$ and $\ket{\phi_n(t)}$, i.e., $I(t)\ket{\phi_n(t)}=\lambda_n\ket{\phi_n(t)}$. $I(t)$ can be used to express an arbitrary solution of the time-dependent Schr{\"o}dinger equation $i\hbar\partial_t\ket{\Psi(t)}=H(t)\ket{\Psi(t)}$ as a superposition of dynamical modes $\ket{\psi_n(t)}$, i.e., $\ket{\Psi(t)}=\sum_nc_n\ket{\psi_n(t)}=\sum_nc_ne^{i\alpha_n(t)}\ket{\phi_n(t)}$ with $c_n$ being time-independent amplitude and $\alpha_n(t)$ being the Lewis-Riesenfeld phase, and the time-dependent unitary evolution operation is 
\begin{equation}
U(t,0)=\sum_{n=1}^{3} e^{i\alpha_{n}(t)}\ket{\phi_{n}(t)}\bra{\phi_{n}(0)}.
\end{equation}

To simplify the calculation of STA, it is convenient to let the system evolves along the one dynamical mode, i.e., 
\begin{equation}
    \ket{\Psi(t)}=\ket{\psi_1(t)}=e^{i\alpha_1(t)}\begin{pmatrix}
        \cos^2\frac{\theta(t)}{2}e^{-i\beta(t)}\\ \frac{1}{\sqrt{2}}\sin\theta(t)\\ \sin^2\frac{\theta(t)}{2}e^{i\beta(t)}
    \end{pmatrix}.
\end{equation}
Thus, by properly designing $\theta(t)$ to satisfy the boundary condition
\begin{eqnarray}\label{Eq:BC1}
\theta(0)=0, \theta(T)=\pi,
\end{eqnarray}
transition $\ket{+1}\to\ket{-1}$ can be achieved. Accordingly, the envelope $\Omega(t)$ can be obtained according to Eq.\ref{Eq:theta}. 

More importantly, the invariant-based calculation of STA  also enables the state transition in the presence of noise. Suppose there is imperfection in biased external magnetic field $B_z$, i.e., $B_z^\prime\to B_z+\delta/(2 \pi \gamma_e)$, which shifts the energy difference between spin state $\ket{+1}$ and $\ket{-1}$ by $2\delta$. In the presence of noisy $B_z^\prime$, the Hamiltonian of NV center is $H^\prime=\Omega(t)J_x/\sqrt{2}+\delta J_z$. $\delta J_z$ can be regarded as a perturbation to $H$, so that the final dynamic mode $\ket{\Psi^\prime(T)}$ after evolution time $T$ can be expanded using Dyson series~\cite{zeng2019PRA,Dridi2020PRL,Barnes2022QST}
\begin{equation}
\begin{split}
\ket{\Psi^\prime(T)}=&\ket{\psi_1(T)}-i\delta\int_0^TdtU(T,t)J_z\ket{\psi_1(t)}\\
&-\delta^2\int_0^Tdt\int_0^tdt^\prime U(T, t)J_zU(t, t^\prime)J_z\ket{\psi_1(t^\prime)}\\
&+\cdots.
\end{split}
\end{equation}
The accuracy of spin transition can be characterized by fidelity between the final dynamic mode $\ket{\Psi^\prime(T)}$ and the desired mode $\ket{\psi_1(T)}$  
\begin{equation}\label{Eq:Fidelity}
\begin{split}
F&=\left|\bra{\psi_1(T)}\Psi^\prime(T)\rangle\right|^2\\
&\simeq 1-\delta^{2}\sum_{n\neq1} \left | \int_{0}^{T} dt\bra{\psi_{1}(t)}J_{z}\ket{\psi_{n}(t)}\right |^{2}.
\end{split}
\end{equation}
It is straightforward that the second term in Eq.~\ref{Eq:Fidelity} can be limited with properly designed $\theta(t)$~($\Omega(t)$).   

\begin{figure*}[htbp]
\includegraphics[width=\linewidth]{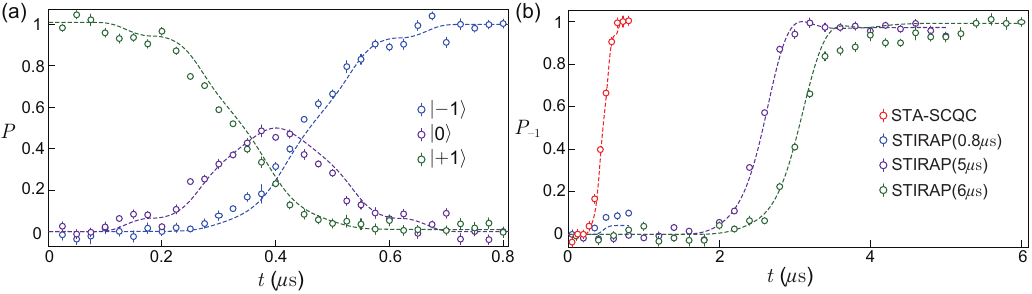}
\caption{\label{Fig:3} (a) Experimental results of $P_0$ and $P_{\pm1}$ with {STA-SCQC}. (b) Experimental results of $P_{-1}$ with {STA-SCQC} and STIRAP. The dashed lines represent the theoretical predictions, and the error bars are the statistical errors. }
\end{figure*}

The direct calculation of $\theta(t)$ is complicated, it is convenient to obtain $\Omega(t)$ from a three-dimensional arc-length function $\bm{r}(t)$~\cite{zeng2019PRA}. To eliminate the second term in Eq.~\ref{Eq:Fidelity}, one can define an auxiliary operator $m(t)$
\begin{equation}
m(t)=\int_{0}^{t} U^\dagger(t^{\prime},0)J_{z}U(t^{\prime},0)dt^{\prime}, 
\end{equation}
which can also be represented with $J_\nu$ by $m(t)=\bm{r}(t)\cdot\bm{J}=x(t)J_{x}+y(t)J_{y}+z(t)J_{z}$ with $\bm{r}(t)$ being a three-dimensional curve. The expansion coefficients $x(t)=\Tr[J_{x}m(t)]/2$, $y(t)=\Tr[J_{y}m(t)]/2$ and $z(t)=\Tr[J_{z}m(t)]/2$ are projective coordinates of $\bm{r}(t)$ in the three dimensional Euclidean space. Thus, if $m(t)$ satisfies the condition $m(T)=0$, the second item in Eq.~\ref{Eq:Fidelity} can be eliminated. The constraint of $m(0)=m(T)=0$ indicates the curve $\bm{r}(t)$ should be closed and $\bm{r}(0)=\bm{r}(T)=0$. The boundary condition in Eq.~\ref{Eq:BC1} impose constraint on the derivative of $\bm{r}(t)$, i.e., 
\begin{equation}\label{Eq:const}
    \dot{\bm{r}}(0)=[0,0,1], \dot{\bm{r}}(T)=[0,0,-1]
\end{equation}
with
\begin{equation}\label{Eq:derir}
\dot{\bm{r}}(t)=[-\sin\theta(t)\cos\alpha_1(t), -\sin\theta(t)\sin\alpha_1(t), \cos\theta(t)].
\end{equation} 
Then, the envelope of $\Omega(t)$ can be determined by the  curvature of $\bm{r}(t)$
\begin{equation}\label{Eq:curvature}
   \Omega(t)=||[H(t),J_z]||=||\ddot{\bm{r}}(t)||,
\end{equation}
where $||\cdot||$ is the Frobenius norm. 

In this work, we propose to construct $\Omega(t)$ according to the spacial curve $\bm{r}(\zeta)$ in form of  \begin{equation}\label{Eq:spacecurve}
\begin{split}
\bm{r}(\zeta) & =[0,y(\zeta),z(\zeta)] \\
&=\left[0,\frac{-3\sqrt{2} \sin(4\pi\zeta)}{4+4\cos^{2}(2\pi\zeta)}f(\zeta), \frac{2\sqrt{2}\sin(2\pi\zeta)}{1+\cos^{2}(2\pi\zeta)}\right]
\end{split}
\end{equation} 
with $\zeta\in[0, 0.5]$. $f(\zeta)=(\int_0^{\mu_\zeta}e^{-l^2}dl-\int_0^{\eta_\zeta}e^{-l^2}dl)/\sqrt{\pi}$ is the modulation function, where
\begin{equation}\label{Eq:parameters}
    \mu_\zeta=\frac{2[-a-2\pi(\zeta-0.5+b)]}{a}, 
    \eta_\zeta=\frac{2[a-2\pi(\zeta-b)]}{a}
\end{equation}
with $a$ and $b$ being constant parameters. Note that $\zeta$ is not a arc-length parameter. $\bm{r}(\zeta)$ can be reparameterized by arc-length function 
\begin{equation}\label{Eq:arclength}
   L(\zeta)=\int_0^\zeta\sqrt{[y^\prime(\zeta)]^2+[z^\prime(\zeta)]^2}d\zeta. 
\end{equation}
By substituting $L(\zeta)$ back to Eq.~\ref{Eq:spacecurve}, we obtain the spacial curve parameterized by arc-length $\bm{r}(L)$. Thus, the envelope of Rabi frequency $\Omega(L)$ can be obtained according to Eq.~\ref{Eq:curvature}. 

We fix the maximal Rabi frequency in $\Omega(L)$ by $\Omega_\text{max}=2\pi\times1.9$~MHz, then optimize the value of  $a$ and $b$ in Eq.~\ref{Eq:parameters} to obtain the minimal $L$~(evolution time $T$). The simulation results are shown in Fig.~\ref{Fig:2}~(a), which implies the {optimized} $T\approx0.8~\mu$s with $a=0.15\pi$ and $b=0.06$~(red dot in Fig.~\ref{Fig:2}~(a)). {Note that the choice of closed space curve is flexible. The optimal $T$ is obtained with constraint of space curve in Eq.~\ref{Eq:spacecurve}. Thus, the envelope of $\Omega(t)$ in Fig.~\ref{Fig:2}~(d) is obtained by substituting the optimized $a$ and $b$ into Eqs.~\ref{Eq:curvature}-~\ref{Eq:arclength}.} We use the pulses sequence shown in Fig.~\ref{Fig:2}~(b) to implement the {STA-SCQC} transition, where the spin state is initialized on $\ket{0}$ with a 5~$\mu$s-long laser pulse and then excited to $\ket{+1}$ with a 300~ns-long MW $\pi$-pulse at frequency of $\omega_+$. The MW pulse to implement {STA-SCQC} transition is generated with the setup shown in Fig.~\ref{Fig:2}~(c), where the MW signals at resonate frequencies $\omega_+$ and $\omega_-$ are generated with a MW generator~(MWG) and MW vector signal generator~(VSG) respectively. The amplitude of MW field $A(t)$ is modulated by an Arbitrary Waveform Generator~(AWG). Two modulated MW fields are combined with a MW combiner, then amplified and sent into the MW antenna~(copper wire). 

We first carefully adjusted MW amplitude $A$ to realize Rabi oscillation between $\ket{0}\leftrightarrow\ket{\pm1}$ with Rabi frequency of $2\pi\times 1.9$~MHz, and the corresponding MW amplitude is denoted as {$A_\text{max}$. As $\Omega$ is proportional to the the MW amplitude $A$,}  $A_\text{max}$ acts as a benchmark to modulate $A(t)$ to realize $\Omega(t)$. After time $t$, we detect the populations of states $\ket{0}$, $\ket{+1}$ and $\ket{-1}$. $\ket{0}$ is readout by a 300~ns-long laser, and $\ket{+1}$~($\ket{-1}$) is readout by applying a $\pi$-pulse~(dashed boxes in Fig.~\ref{Fig:2}~(b)) at frequency $\omega_+$~($\omega_-$) before the readout laser. The results of $P_0$, $P_{+1}$ and $P_{-1}$ are shown in Fig.~\ref{Fig:3}~(a), the transfer of spin states from $\ket{+1}$ to $\ket{-1}$ is achieved with high efficiency~($99\%$)~at $t\approx0.8~\mu$s, which agrees very well with simulations. The speedup of {STA-SCQC} is confirmed by comparison with spin transition with STIRAP protocol, in which two MW pulses with Gaussian shape $A_{+}(t)=A_{+} \exp[-(t-T/2+\Delta\tau)^{2}/\sigma ^{2}]$ and $A_{-}(t)=A_{-}\exp[-(t-T/2-\Delta\tau)^{2}/\sigma ^{2}]$~(as shown in Fig.~\ref{Fig:2}(e)) are applied partly overlapping and in a counter-intuitive pulse order. $A _{+}(t)$ and $A _{-}(t)$ are also referred as Stokes pulse and pumping pulse respectively. $2\sigma=T/3$ is the full-width at half-maximal of the pulse and $2\Delta\tau=T/5$ is the separation of time between  $A_{+}(t)$ and $A_{-}(t)$. We set $A_{+}=A_{-}=A_\text{max}$ and the operation time $T=0.8~\mu$s, $T=5~\mu$s and $T=6~\mu$s respectively, and the results of $P_{-1}$ are shown in Fig.~\ref{Fig:3}~(b). We observe that spin transition can not be achieved in operation time $T=0.8~\mu$s as the adiabaticity criteria $2\Omega_\text{max}\sigma\gg 1$ is not satisfied~\cite{Vitanov2001}. By increasing the operation time $T$ up to 5~$\mu$s, the adiabaticity criteria is fulfilled and the spin transition is achieved. However, the time cost of STIRAP is at least six times longer than that of {STA-SCQC}.  

\begin{figure*}[ht!bp]
\includegraphics[width=\linewidth]{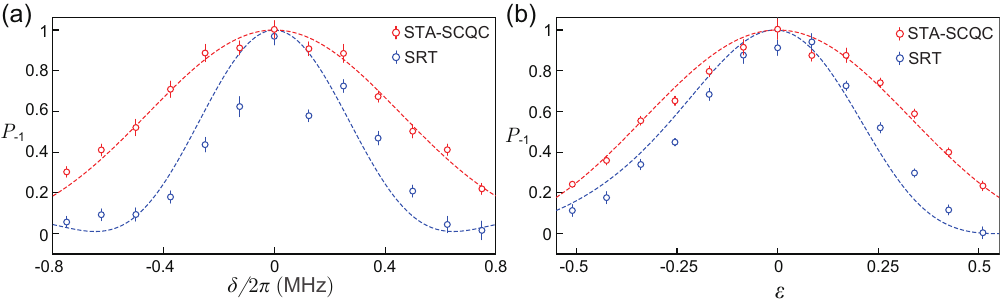}
\caption{\label{Fig:4}(a) Experimental results of $P_{-1}$ with {STA-SCQC}~(red circles) and SRT~(blue circles) under the fluctuation of $\delta$. (b) Experimental results of $P_{-1}$ with {STA-SCQC}~(red circles) and SRT~(blue circles) under the fluctuation of $\epsilon$.}
\end{figure*} 

Next, we investigate the robustness of {STA-SCQC} in the presence of fluctuation in biased magnet field $B_z$. Experimentally, the fluctuation of $B_z$ is equivalently realized by setting frequencies of MW fields as $\omega_+^\prime=\omega_++\delta/2 \pi$ and $\omega_-^\prime=\omega_--\delta/2 \pi$. To give a comparison, we also carry out the spin transition with SRT. {SRT enables population transfer between $\ket{+1}$ and $\ket{-1}$ with the assistance of $\ket{0}$ using a pair of squared pulses as show in Fig.~\ref{Fig:2}~(f), where the corresponding frequency detuning is $\Delta$ with respect to $\ket{0}$}. In our experiment, we set $A_+=A_-=A_\text{max}$, $T=0.8~\mu$s and $\omega_{\pm,\Delta}=\omega_{\pm}-\Delta$ with $\Delta=2.5$~MHz. In a similar approach, the errors are introduced by setting $\omega_{+,\Delta}^\prime=\omega_{+,\Delta}+\delta/2 \pi$ and $\omega_{-,\Delta}^\prime=\omega_{-,\Delta}-\delta/2 \pi$. The results of $P_{-1}$ with noisy {STA-SCQC} and SRT pulses are shown with red and blue circles in Fig.~\ref{Fig:4}~(a). We observe that both {STA-SCQC} and SRT can achieve high-fidelity spin transition at $\delta=0$.  As $|\delta|$ increases, $P_{-1}$ with SRT decreases faster than that with {STA-SCQC}. The transfer efficiency with optimized STA is larger than 80\% when $|\delta|\leq2 \pi\times 300$~kHz, which corresponds to fluctuation of 10.7~$\mu$T in $B_z$. Interestingly, this is also the same scale of nuclear spin bath noise~\cite{Abobeih2019Nature}. The robustness of {STA-SCQC} is also investigated under the fluctuation of the amplitude of MW field $A(t)$, which would introduce the errors in Rabi frequency, i.e., $\Omega(t)\to(1+\epsilon)\Omega(t)$. The results of $P_{-1}$ are shown in Fig.~\ref{Fig:4}~(b). Although $\Omega(t)$ with {STA-SCQC} is not specifically designed for this kind of imperfection, it is more robust than SRT.

In conclusion, we demonstrate STA spin transition between ground states of a single NV center, where the shortcut is designed with the assistance of invariant-based inverse engineering and {SCQC}. The experimental results show the speedup of spin transition and robustness against experimental imperfections, compared to the traditional Raman control schemes. Very recently, a jump protocol to speedup the adiabatic process has been proposed~\cite{liu2022arXiv} and implemented to speedup spin transition in NV center system~\cite{gong2023PRA}, where two MW field $A_{+}(t)$ and $A_{-}(t)$ should be individually modulated. In our demonstration, identical modulation of two MW field is required, which alleviates the experimental complexity. {In contrast other advanced STIRAP-like processes focusing on energy leakage on exited state and noise in the amplitude of MW field~\cite{Laforgue2019PRA,Laforgue2020PRA,Liu2023entropy}, our demonstration shows the robustness against the detuning errors in driven MW field, which is equivalent to the drifts of the bias magnetic field $B_z$ applied on NV center. The drifts of $B_z$ vary randomly and are challenging to calibrate in experiment~\cite{rong2014PRL}.} Our results benefit quantum information processing with NV center platform, in which the full control over ground states is required.

\bigskip

\noindent{\emph{Data Availability Statements.---}}
The data are available from the corresponding author on reasonable request.

\begin{acknowledgments}
\noindent{\emph{Acknowledgments.---}}
We thank K.~Z. Li for the insightful discussion on this work. This work is supported by Taishan Scholar of Shandong Province (Grant No.~tsqn202103013).
\end{acknowledgments}

\bibliography{STA}

\end{document}